\documentclass[aip,apl,preprint]{revtex4-1}
\usepackage{graphicx}
\usepackage{subfigure}

\begin{document}
\title{Graphene quantum dots embedded in hexagonal boron nitride sheets} 
\author{Junwen Li}
\author{Vivek B. Shenoy}
\email{vivek\_shenoy@brown.edu}
\affiliation{School of Engineering, Brown University, Providence, RI 02912, USA}

\date{\today}

\begin{abstract}
We have carried out first-principles calculations on electronic properties of graphene quantum dots embedded in hexagonal boron nitride monolayer sheets. The calculations with density functional theory show that the band gaps of quantum dots are determined by the quantum confinement effects and the hybridization of $\pi$ orbitals from B, N and C atoms. The energy states near the Fermi level are found to be strongly localized within and in the vicinity of the quantum dots.  
\end{abstract}


\maketitle 

Graphene has attracted a great deal of interest since it was exfoliated from bulk graphite.\cite{Novoselov_science_2004, Novoselov_pnas_2005}  Because graphene has a number of unusual properties such as massless fermions and high carrier mobility, graphene materials are expected to play an important role in fundamental research and in future electronic device applications.\cite{Novoselov_nature_2005, Lin_science_2010} Two-dimensional graphene has zero band gap with linear energy dispersion near the Fermi level. Many efforts have been devoted to opening a band gap in order to fit various needs. One promising method is to confine charge carriers in two dimensions by cutting graphene into ribbons,\cite{Han_prl_2007,Delgado_nl_2008,Jia_science_2009, Jiao_nature_2009} of which the zigzag graphene nanoribbons are of special interest because of the spin-polarized edge states predicted to be half-metallic upon applying transverse electric field.\cite{Son_nature_2006} 

Electronic states in graphene could be further confined in the basal plane to make quantum dots (QDs) and there are already a few investigations on this idea.  Zhang \textit{et al.}\cite{Zhang_jpcm_2010} showed that the surface corrugation of graphene on Ru(0001) can form QDs. Ponomarenko \textit{et al.}\cite{Ponomarenko_science_2008} carved out graphene dot devices behaving as single-electron transistors and exhibiting coulomb blockade behavior. 
Yan \textit{et al.}\cite{Yan_nl_2010} found that solution-processable graphene dot can be used as light absorbers for photovoltaic applications.  The QDs also could be hosted in other two-dimensional systems. Singh \textit{et al.}\cite{Singh_acsnano_2010} using density functional theory and the tight-binding method, proposed that H-vacancy clusters in graphane could form graphene quantum dots having size dependence of energy gaps typical for confined Dirac fermions.

Along with the development of graphene, rapid progress has been made on the study of boron nitride materials. Recently graphene-like hexagonal boron nitride (h-BN) sheets have been fabricated experimentally.\cite{Pacele_apl_2008,Han_apl_2008,Jin_prl_2009,Alem_prb_2009, Song_nl_2010, Muller_prb_2010} When C atoms mix with boron nitride, many interesting properties can emerge. Even without the need of externally applying electric field, hybrid C-BN nanotubes and  nanoribbons could exhibit intrinsic half-metallicity\cite{Dutta_prl_2009, Huang_apl_2010,Pruneda_prb_2010,Ding_apl_2009} and the C doping can induce magnetism in BN nanotubes.\cite{Wu_apl_2005}

In the present letter, we report the first-principles calculations on the graphene QDs with h-BN sheets as host materials. Our results show that the hybridization between $\pi$ orbitals of B, N and C atoms determines the energy gap together with the quantum confinement effect. The states close to Fermi level $E_F$ are found strongly localized around the C-doped region. 

The calculations were performed with spin-unrestricted density functional theory as implemented in the SIESTA code.\cite{Soler_jpcm_2002} The norm-conserving Troullier-Martins pseudopotentials were used to model the interaction between ionic cores and valence electrons and double-$\zeta$ basis set plus polarization orbitals were employed to construct the valence orbitals. The exchange-correlation functional was described with Perdew-Burke-Ernzerhof approximation. A grid cut-off of 200 Ry was used. The graphene QDs embedded in h-BN monolayer are simulated by doping C atoms in a h-BN sheet of rectangular shape. A vacuum region of 16.5~\AA\ was added along the direction normal to the sheet plane to avoid the interaction between periodic images of supercells. Relaxations were carried out on all structures investigated until the force on each atom is less than 0.04 eV/\AA. The k-points sampling schemes of 1x1x1 and 2x2x1  were employed for geometry optimization and density of states (DOS) analysis, respectively.

\begin{figure}[h]
  \includegraphics[scale=1]{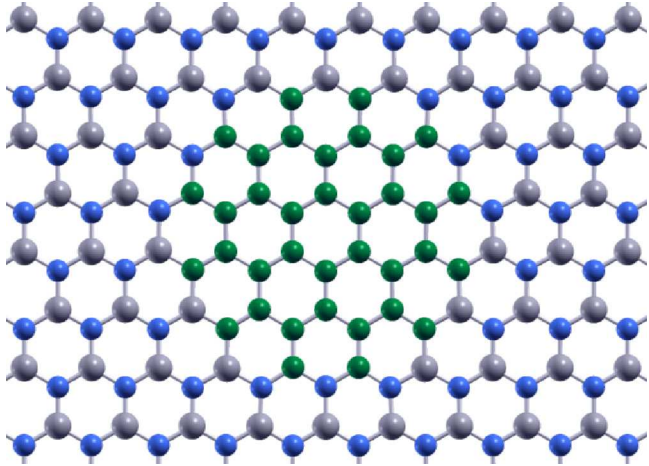}
  \caption{Structural model of QD C$_{36}$ embedded in h-BN sheet. B, N and C atoms are represented by gray, blue and dark-green spheres, respectively.}
  \label{dot_geom}
\end{figure}

The QDs of high symmetry, as depicted in Fig.~\ref{dot_geom}, were chosen in this study. They were constructed by replacing all B and N atoms enclosed within a circle of diameter $d$ by C atoms with the total number of atoms fixed at 512. We varied $d$ from 2.85~\AA\ to 12.55~\AA, corresponding to the total C atoms $n$ = 6, 12, 24, 36 and 54, to model QDs of different sizes denoted by C$_n$. 

\begin{figure}[h]
  \includegraphics[scale=1]{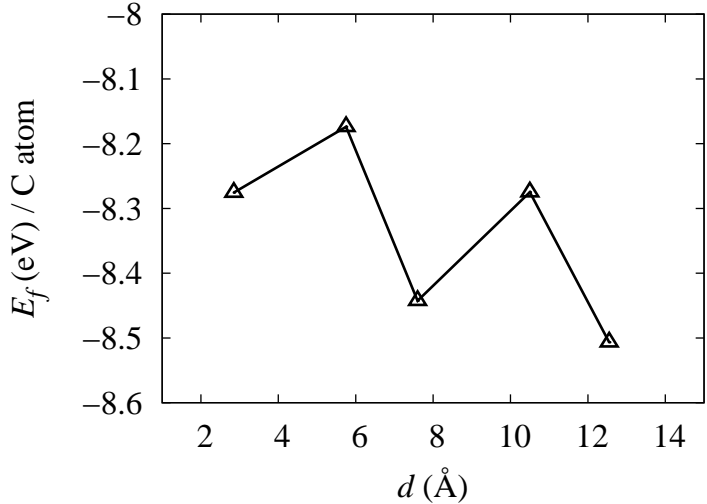}
  \caption{Formation energy $E_f$ as a function of QDs diameter $d$.}
  \label{dot_formation}
\end{figure}

The stability of QDs is estimated from the formation energy which can be expressed as
\begin{displaymath}
 E_f = E_{\mbox{\small{tot}}} - m\mu_{\mbox{\small{BN}}} + 2m\mu_{\mbox{\small{C}}} {\rm ,}
\end{displaymath}
 where $E_{\mbox{\small{tot}}}$ is the total energy of QDs and $\mu_{\mbox{\small{BN}}}$ and  $\mu_{\mbox{\small{C}}}$ are the chemical potentials of BN pair and C atom, respectively. $\mu_{\mbox{\small{BN}}}$ and $\mu_{\mbox{\small{C}}}$ are calculated from the cohesive energies of a BN pair in an infinite h-BN monolayer and of each C atom in a single graphene sheet, respectively. Here $m$ represents the total number of BN pairs replaced by C atoms. We display the formation energies per C atom as a function of QDs size in Fig.~\ref{dot_formation}. Roughly, the formation energies decrease with increasing quantum dot size. The detailed size dependence, however, is oscillating instead of monotonic. Note that the QDs C$_6$, C$_{24}$, and C$_{54}$ comprised of complete aromatic rings have lower formation energy than C$_{12}$ and C$_{36}$. This shape preference was also observed in graphene nanoribbons embedded in boron nitride sheets\cite{Ding_apl_2009} and graphene QDs originating from H-vacancies in graphane.\cite{Singh_acsnano_2010}

\begin{figure}[h]
  \includegraphics[scale=1]{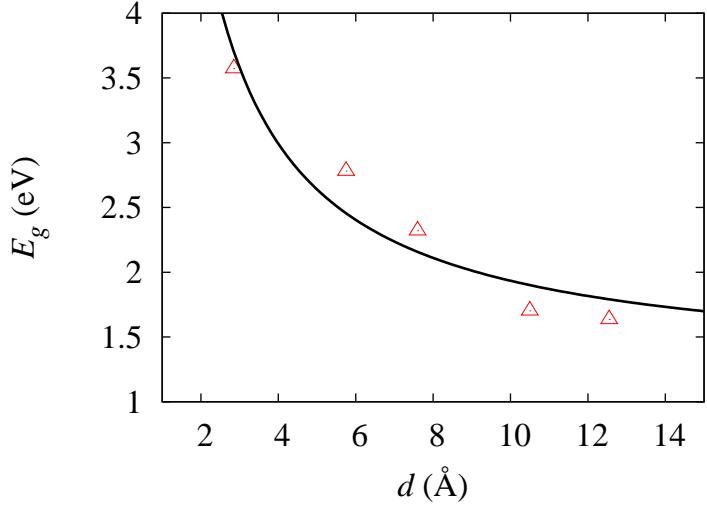}
  \caption{Energy gap $E_g$ as a function of QD diameter $d$. Solid cure is drawn as a guide to the eyes.}
  \label{dot_gap}
\end{figure}

For pure h-BN sheet, our calculated energy gap is 4.57 eV comparable to 4.61 eV reported from calculations with plane wave basis.\cite{Sahin_prb_2009} After pairs of B and N were substituted by C, the energy gap is reduced. Fig.~\ref{dot_gap} depicts the change in energy gap for various values of $d$. When increasing the QDs size, the energy gap reduces from 3.57 eV to 1.64 eV. In isolated graphene flakes terminated with H atoms, the gap change with size is mainly determined by quantum confinement effect since H atom only has one $s$ electron forming $sp^2$ $\sigma$ bonds far away from the Fermi level. This is similar to the QDs from graphane except the $sp^3$ $\sigma$ bonds are formed instead. When the QDs are surrounded by B and N atoms in the present study, the situation is different. In addition to the confinement effect, the 2$p$ electrons of B and N will form $\pi$ bonding near the Fermi level with 2$p$ electrons of C. We can see this point based on the DOS analysis.

\begin{figure}[]
  \includegraphics[scale=1,angle=0]{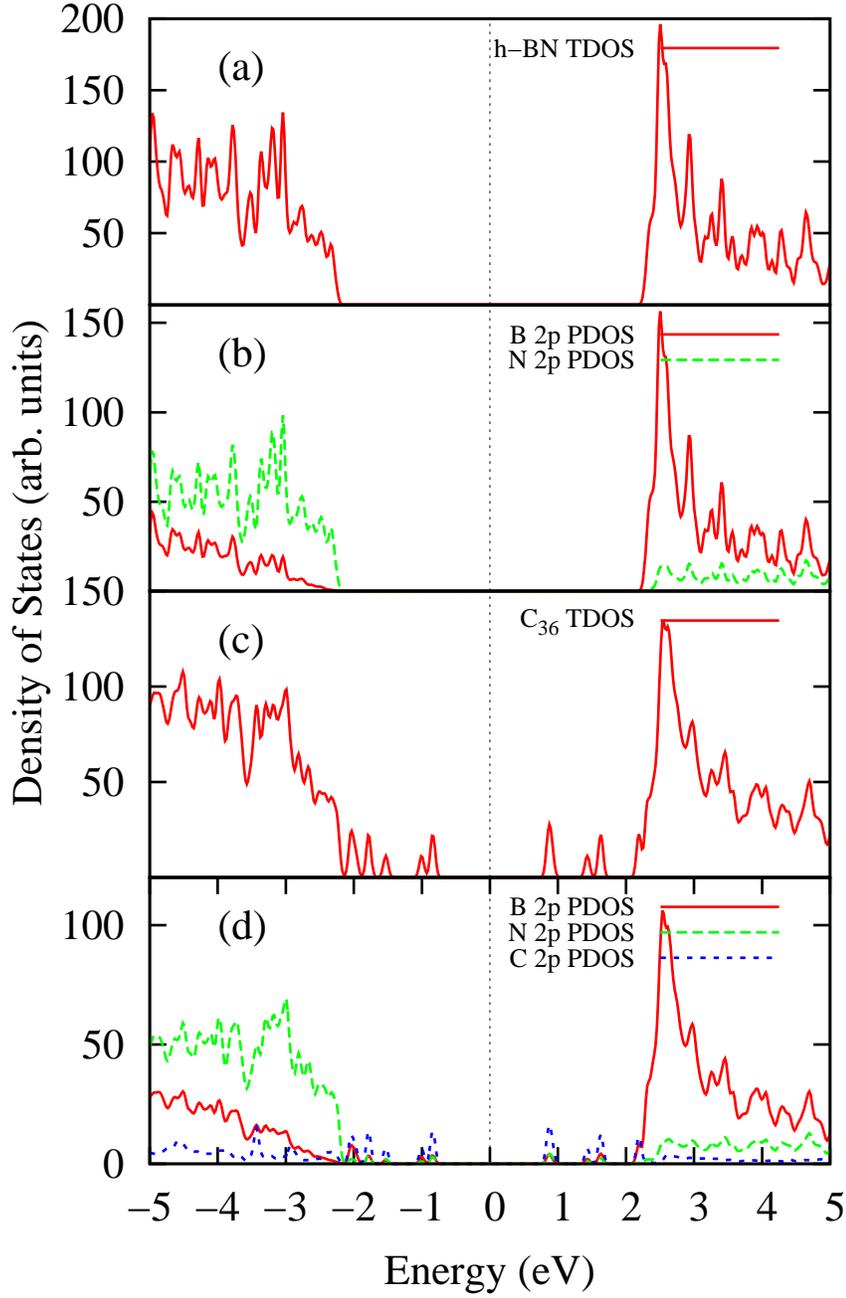}
  \caption{(a) TDOS and (b) PDOS of h-BN sheet. (c) TDOS and (d) PDOS of  QD C$_{36}$.  PDOS represent 2$p$ orbital contributions of B, N and C as indicated. The vertical dotted line indicates the Fermi level as zero energy reference. We apply a broadening of 50 meV to the energy levels.}
  \label{dot_pdos}
\end{figure}

\begin{figure}[]
  \subfigure[]{
    \includegraphics[scale=1]{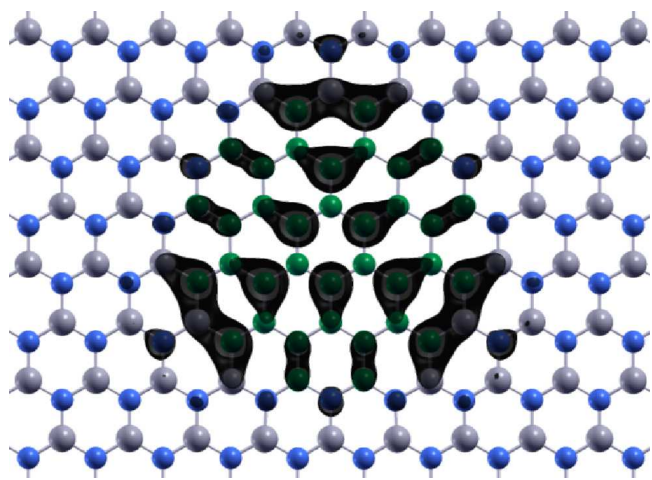}
    \label{ldos_homo}
  }
  \subfigure[]{
    \includegraphics[scale=1]{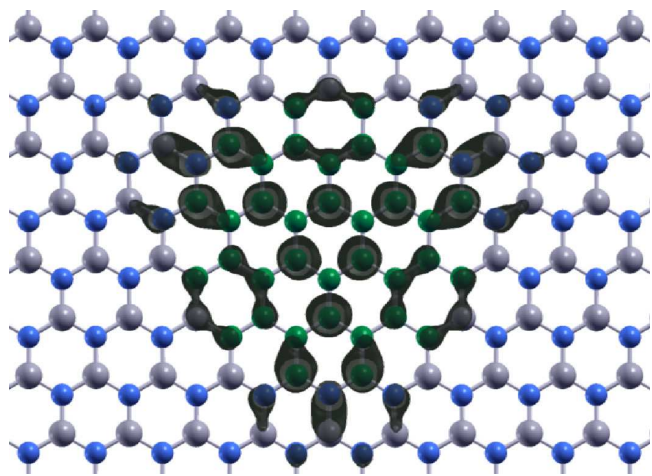}
    \label{ldos_lumo}
  }
  \caption{Isosurfaces of LDOS of (a) the HOMO and (b) the LUMO for QD C$_{36}$. The isovalue is 10\% of the maximum.}
    
\end{figure}

In Fig.~\ref{dot_pdos} we show the total density of states (TDOS) and partial density of states (PDOS) decomposed onto 2$p$ orbitals for h-BN sheet and QD C$_{36}$ ($d$ = $10.50$~\AA). For pure h-BN sheet, the top of valence bands are dominated by 2$p$ orbitals from N  and the bottom of conduction bands mainly originate from 2$p$ orbitals of B atoms. The introduction of graphene QDs has almost no effect in the energy range of valence and conduction bands of h-BN sheet. However, the near-$E_F$ DOS are largely modified and several peaks show up. The peaks 0.85 eV below and above $E_F$ are almost dispersionless and there is also strong orbital hybridization between 2$p$ orbitals of B, N and C atoms. The mixing of 2$p$ orbitals will form bonding and anti-bonding states. The energies of the 2$p$ orbital are different with  B higher than  N and C in the middle. So near the Fermi level, only $\pi_{\mbox{\tiny{C-B}}}$ and $\pi^*_{\mbox{\tiny{C-N}}}$ are the relevant bands.  To see how these states distribute in real space, we show the local density of states (LDOS) for the highest occupied molecular orbitals (HOMO) in Fig.~\ref{ldos_homo} and for the lowest unoccupied molecular orbitals (LUMO) in Fig.~\ref{ldos_lumo}. These energy states are strongly localized in the C region and also have considerable distribution on B and N atoms at the interface of C-BN in the form of $\pi_{\mbox{\tiny{C-B}}}$ and $\pi^*_{\mbox{\tiny{C-N}}}$ states. In the interior region, compared to HOMO, the LUMO has more nodes which increase the kinetic contribution in energy. The discussion based on C$_{36}$ also applies on other QDs in which we observed similar orbital hybridization and carrier localization. 

\clearpage
In summary, we have performed pseudopential calculations with numerical atomic orbitals basis to investigate the electronic properties of graphene QDs embedded in h-BN sheet. Our results show that the orbital hybridization between 2$p$ orbitals of B, N and C and the quantum confinement together determine the energy gaps of QDs. The dispersionless energy states near Fermi level are strongly confined in the QDs region. Great progress in the synthesis of single layer h-BN sheets and graphene-BN composites\cite{Ci_nmat_2010} makes the kind of quantum dots discussed in our work potentially promising candidates in electronic and optoelectronic applications.

We gratefully acknowledge research support from the NSF and NRI through the Brown University MRSEC program and the NSF through the grants CMMI-0825771, DMS-0914648 and DMS-0854919.

\end{document}